**Visualizing Nodal Heavy Fermion Superconductivity in CeCoIn$_5$**

Brian B. Zhou[1*], Shashank Misra[1*], Eduardo H. da Silva Neto[1], Pegor Aynajian[1], Ryan E. Baumbach[2], J. D. Thompson[2], Eric D. Bauer[2], and Ali Yazdani[1]

[1]*Joseph Henry Laboratories & Department of Physics, Princeton University, Princeton, New Jersey 08544, USA*

[2]*Condensed Matter and Magnet Science, Los Alamos National Laboratory, Los Alamos, New Mexico 87545, USA*

* These authors contributed equally to this work.

**Understanding the origin of superconductivity in strongly correlated electron systems continues to be at the forefront of unsolved problems in all of physics.[1] Among the heavy *f*-electron systems, CeCoIn$_5$ is one of the most fascinating, as it shares many of the characteristics of correlated *d*-electron high-$T_c$ cuprate and pnictide superconductors[2-4], including the competition between antiferromagnetism and superconductivity.[5] While there has been evidence for unconventional pairing in this compound[6-11], high-resolution spectroscopic measurements of the superconducting state have been lacking. Previously, we have used high-resolution scanning tunneling microscopy (STM) techniques to visualize the emergence of heavy-fermion excitations in CeCoIn$_5$ and demonstrate the composite nature of these excitations well above $T_c$.[12] Here we extend these techniques to much lower temperatures to investigate how superconductivity develops within a strongly correlated band of composite excitations. We find the spectrum of heavy excitations to be strongly modified just prior to the onset of superconductivity by a suppression of the spectral weight near the Fermi energy**

**($E_F$), reminiscent of the pseudogap state[13, 14] in the cuprates. By measuring the response of superconductivity to various perturbations, through both quasiparticle interference and local pair-breaking experiments, we demonstrate the nodal *d*-wave character of superconducting pairing in CeCoIn$_5$.**

CeCoIn$_5$ undergoes a superconducting transition at 2.3 K. Despite evidence of unconventional pairing, consensus on the mechanism of pairing and direct experimental verification of the order parameter symmetry are still lacking.[6-9, 11] Moreover, experiments have suggested that superconductivity in this compound emerges from a state of unconventional quasiparticle excitations with a pseudogap phase similar to that found in underdoped high-T$_c$ cuprates.[15-17] Previously, we demonstrated that STM spectroscopic techniques can be used to directly visualize the emergence of heavy fermion excitations in CeCoIn$_5$ and their quantum critical nature.[12] Through these measurements, we also demonstrated the composite nature of heavy quasiparticles and showed their band formation as the *f*-electrons hybridize with the *spd*-electrons starting at 70 K, well above T$_c$.[12] This previous breakthrough, together with our recent development of a high-resolution milli-Kelvin STM, offers a unique opportunity to measure how superconductivity emerges in a heavy electron system.

Figure 1 shows STM topographs of the two commonly observed atomically ordered surfaces of CeCoIn$_5$ produced after the cleaving of single crystals *in situ* in the ultra-high vacuum environment of our milli-Kelvin STM. We have previously shown through experiments and theoretical modeling that different surface terminations change the coupling between the tunneling electrons and the composite heavy fermion excitations in this compound.[12] Tunneling into such composite states can be influenced not only by the coupling of the tip to *spd*- or *f*-like component of such states but also by the interference between these two tunneling processes. On surface A, tunneling measurements are more sensitive to the lighter component of the composite band structure, and accordingly, the spectra show evidence for a hybridization gap centered at +9 mV, as shown in Fig. 1d. At temperatures below $T_c$, this hybridization gap is modified by the onset of an energy gap associated with superconductivity (Fig. 1c,d), as further confirmed by its suppression with the application of a magnetic field larger than the bulk upper critical field ($H_{C2}$ = 5.0 T perpendicular to the basal plane of this tetragonal system) of $CeCoIn_5$ (see supplementary section I).

Instead of focusing on measurements of surface A, where the tunneling is dominated by the lighter part of the composite band, we turn to measurements of surface B. On this surface tunneling directly probes narrow bands of heavy excitations which result in a peak in the density of states near $E_F$ (Fig. 1e). Lowering the temperature from 7.2 K to 5.3 K, above $T_c$, we find that this peak is modified by the onset of a pseudogap-like feature at a smaller energy scale. Further cooling shows the onset of a distinct superconducting gap below $T_c$ inside the pseudogap. Measurements in a magnetic field corroborate our finding that the lowest energy scale on surface B (~ ±500 μV, as shown in Fig. 1c) is indeed associated with pairing, as it disappears above $H_{C2}$, while the intermediate energy scale pseudogap remains present at low temperature in the absence of superconductivity at high magnetic field (Fig. 1f). This behavior is reminiscent of the pseudogap found in underdoped cuprates, where the superconducting gap opens inside an energy scale describing strong correlations that onset above $T_c$. However, unlike cuprates, here we clearly distinguish between the two energy scales by performing high-resolution spectroscopy in a magnetic field large enough to fully suppress superconductivity. Detailed measurements of changes in the spectra with the magnetic field also confirm that the transition out of the superconducting state at $H_{C2}$ is first order (see supplementary section I), showing that our measurements are consistent with the bulk phase diagram of CeCoIn$_5$.

The spectroscopic measurements suggest that electronic or magnetic correlations alter the spectrum of heavy excitations by producing a pseudogap within which pairing takes place. These measurements also show the shapes of the spectra at the lowest temperature to be most consistent with a $d$-wave superconducting gap, as they have a nearly linear density of states near zero energy (Fig. 1c). However, measurements on all surfaces and on several samples reveal that this $d$-wave gap (with a magnitude of 535 ± 35 μV, consistent with that extracted from point contact data[18, 19]) is filled (40%) with low energy excitations—a feature that cannot be explained by simple thermal broadening (determined to be 245 mK from measurements on a single-crystal Al sample, see supplementary section II). The complex multiband structure of CeCoIn$_5$ could involve different gaps on different Fermi surface sheets, and there is the possibility that some remain ungapped even at temperatures well below $T_c$.[20] Another contribution to the in-gap density of states could come from surface impurities, since even non-magnetic impurities perturb a nodal superconductor, as we demonstrate below. Before we address the nature of the in-gap excitations, we first demonstrate in more detail the connection between pairing and the heavy fermionic states of CeCoIn$_5$.

Energy-resolved spectroscopic mapping with the STM can be used to measure the interference of quasiparticles (QPI) in order to examine the heavy Fermi surface. As shown in Figure 2a-d, features in the discrete Fourier transform (DFT) of these maps show wavevectors related to the elastic momentum transfer $Q(E)$, connecting the initial and the final momentum states on the contours of constant energy. Previous theoretical calculations, quantum oscillation, and angle resolved photoemission spectroscopy measurements have shown CeCoIn$_5$ to have a complex three-dimensional band structure, with the $\alpha$ and $\beta$ bands being the most relevant near $E_F$ (Fig. 2e).[21-23] Our previous QPI measurements on surface A show features that are most consistent with $2k_F$ scattering originating from the $\alpha$ band. The QPI measurements presented here on surface B display scattering wavevectors originating from a larger Fermi surface volume and are more consistent with scattering involving the $\beta$ band (see supplementary section IV). Since QPI does not probe the Fermi surface directly, inferring a unique Fermi surface in a three-dimensional, multi-band material without making large number of assumptions is not possible (see supplementary section V). Nevertheless, the results of QPI measurements (Fig. 2a-d) together with spectroscopic measurements (Fig. 1e) demonstrate that the superconducting instability occurs within a correlated heavy quasiparticle band of CeCoIn$_5$ with a large density of states at the Fermi energy.

We focus our discussion next on the momentum structure of the superconducting gap, first by examining the conductance maps in this energy window on the same area of the sample (with the same tip) in the normal ($H > H_{C2}$) and superconducting ($H = 0$) states of CeCoIn$_5$. As the data in Fig. 2f-o demonstrate, we observe clear differences between the DFT maps in the superconducting *(H = 0)* and normal states *(H = 5.7 T)*. Typically, quasiparticle interference at low energies in a superconductor is associated with the scattering of Bogoliubov-de Gennes (BdG) excitations and is often analyzed to obtain information about the momentum structure of the superconducting gap.[24-26] In particular, contrasting the zero-energy DFTs in the superconducting (Fig. 2h) and normal (Fig. 2m) states, we see an enhancement of quasiparticle interference at wavevector $Q_3$ (see also supplementary section VII), suggestive of nodal BdG quasiparticles in a *d*-wave superconductor. However, if such features were only due to BdG-QPI, then they should display a particle-hole symmetric dispersion in their energy-momentum structure away from the nodes, as seen for example in similar measurements of high-$T_c$ cuprates.[24] The absence of such particle-hole symmetry in our data (Fig. 2f-j) together with the large zero-bias density of states (40%, see Fig. 1c) suggests that such QPI measurements are complicated by an ungapped portion of the Fermi surface or by in-gap impurity-induced states, which are expected to have a particle-hole asymmetric structure (see measurements & discussion below). These complications together with complex three-dimensional nature of the Fermi surface of this compound makes extraction of the gap function from such QPI measurements unreliable (see supplementary section VI).

In contrast, using the power of STM to probe the real space structure of electronic states, it is still possible to find direct spatial signatures of the nodal character of superconductivity in CeCoIn$_5$ that do not require multi-parameter modeling or *ad hoc* assumptions to interpret. The first such signature can be found by examining the response of low-energy excitations to extended potential defects such as atomic step edges. Spectroscopic mapping with the STM upon approaching such steps shows direct evidence for the suppression of superconductivity in their immediate vicinity (Fig. 3a-b). This suppression is consistent with the expected response of a nodal superconductor to non-magnetic scattering (Fig. 3c), analogous to similar observations in the cuprates[27], and in marked contrast with our step-edge measurements of the conventional s-wave superconductor Al (see supplementary section II). The data in Fig. 3d provide a direct measure of the Bardeen-Cooper-Schriefer (BCS) coherence length $\xi_{BCS}$ = 56 ± 10 Å, in agreement with $\xi_{BCS} \sim \frac{\hbar v_F}{\pi \Delta} \sim 60$ Å using the gap observed in Fig. 1 (0.5 meV) and the Fermi velocity extracted from Fig. 2 (1.5 x 10$^6$ cm/s).[28]

Application of a magnetic field can also be used to probe the local suppression of heavy-fermion superconductivity in CeCoIn$_5$ due to the presence of vortices and the Abrikosov lattice. As shown in Fig. 4a-b, STM conductance maps can be used to directly visualize the vortex lattice in this compound, which can have different structures depending on the magnetic field. Such structural changes of the vortex lattice (transition between rhombic and square lattices) have been previously studied in neutron scattering experiments[29] and various theoretical models.[30] Complementing these efforts, the STM can be used to probe the electronic states within the vortex core directly, as shown in Fig. 4d, to demonstrate the presence of a zero-energy vortex bound state. Analysis of this core state demonstrates the anisotropic decay of the vortex bound state (Fig. 4c,e and see supplementary section VIII), the angular average (Fig. 4e) of which determines the Ginzburg-Landau coherence length scale ($\xi_{GL}$ = 48 ± 4 Å), consistent with an independent estimate from $\frac{dH_{c2}}{dT}\big|_{T=T_c}$.[31] While observation of such anisotropy is consistent with the nodal character of pairing, an understanding of the role of the underlying Fermi surface symmetry and vortex-vortex interactions is required to model the STM data in more detail.

A more spectacular demonstration of the nodal pairing character in CeCoIn$_5$ can be obtained from examining the spatial structure of in-gap states associated with defects on the surface of cleaved samples. The spatial structure of impurity quasi-bound states, which are mixtures of electron-like and hole-like states, can be a direct probe of the order parameter symmetry.[28, 32] Figure 5 shows an extended defect with a four-fold symmetric structure, which perturbs the low energy excitations of CeCoIn$_5$ by inducing an in-gap state. Probing the spatial structure of these impurity states, we not only find their expected electron-hole asymmetry, but also find that their orientation is consistent with that predicted for a $d_{x2-y2}$ superconductor (Fig. 5b-e and supplementary section IX).[32] The minima (maxima) in the oscillations for hole-like (electron-like) states identify the nodes of the d-wave order to occur at 45° to the atomic axes (Fig. 5h). In fact, these features in the STM conductance maps are identical to those associated with Ni impurities in high-$T_c$ cuprates.[28, 33] However, in contrast to measurements in the cuprates, we are able to determine the spatial structure that such impurities induce on the normal state by suppressing pairing at high magnetic fields. Such measurements allow us to exclude the influences of the normal state band structure, of the impurity shape, or of the tunneling matrix element[28] on the spatial symmetries of the impurity bound state in the superconducting state. Contrasting such measurements for $H > H_{C2}$ (in Fig. 5f-g) with measurements on the same impurity for $H = 0$ (Fig. 5d-e) we directly visualize how nodal superconductivity in CeCoIn$_5$ breaks the symmetry of the normal electronic states in the vicinity of a single atomic defect.

The appearance of a pseudogap and the direct evidence for $d_{x2-y2}$ superconductivity reported here together with previous observations of the competition between anti-ferromagnetism and superconductivity closely ties the phenomenology of the Ce-115 system to that of the high-temperature cuprate superconductors. An important next step in extending this phenomenology would be to explore how the competition between anti-ferromagnetism and superconductivity manifests itself on the atomic scale in STM measurements. Similarly, extending our studies of the electronic structure in magnetic vortices could be used to examine the competition between different types of ordering in the mixed state, and the possible development of the Fulde-Ferrell-Larkin-Ovchinnikov state in this Pauli-limited superconductor.[29, 34, 35]

**Methods**

The single crystal samples (1.5 mm x 1.0 mm x 0.2 mm) used for this study were grown from excess indium at Los Alamos National Laboratory, and were then cleaved along the **c** axis in ultra-high vacuum at room temperature before performing the STM measurements. All data shown from surface B were taken on an undoped sample of $CeCoIn_5$; all surface A measurements were performed on a sample with an effective doping of 0.15% Hg. Bulk transport properties of both samples are indistinguishable. Conductance measurements were made using standard ac lock-in techniques with bias applied to the sample, and were reproduced on different large, atomically flat areas of the sample, having different defect concentrations, and under multiple tunneling conditions ranging up to two orders of magnitude in setpoint current.

We thank K. D. Eaton for helpful discussions. The work at Princeton was primarily supported by grant from DOE-BES. The instrumentation and infrastructure at the Princeton Nanoscale Microscopy Laboratory used for this work were also supported by grants from NSF-DMR1104612, NSF-MRSEC program through Princeton Center for Complex Materials (DMR-0819860), the Linda and Eric Schmidt Transformative Fund, and the W. M. Keck Foundation. Work at Los Alamos was performed under the auspices of the U.S. Department of Energy, Office of Basic Energy Sciences, Division of Materials Science and Engineering.

Author Contributions: B.B.Z. and S.M. performed the STM measurements. B.B.Z., S.M., P.A. and E.H.d.S.N. performed analysis and modelling. R.E.B., J.D.T., and E.D.B. synthesized and characterized the materials. A.Y., B.B.Z , S.M., P.A. and E.H.d.S.N. wrote the manuscript. All authors commented on the manuscript.

Correspondence and requests for materials should be addressed to A.Y. (yazdani@princeton.edu).

**Figure captions:**

**Fig.1: Hybridization, pseudogap, and superconductivity on different surfaces of CeCoIn$_5$.** Topographic image with a set point bias $V$ = -100 mV and current $I$ = 100 pA measured on surface A (a) and with $V$ = -6 mV and $I$ = 100 pA on surface B (b) of CeCoIn$_5$ at 245 mK. Insets in (a) and (b) zoom in on 12x12 nm$^2$ regions on their respective surfaces. The arrows in the figure indicate the in-plane crystallographic *a* and *b* directions. (d,e) Corresponding conductance spectra *G(V)*, proportional to the local electronic density of states on surface A and B carried out at temperatures above and below T$_c$, showing the evolution of the different energy scales ($\Delta_{HG}$: hybridization gap; $\Delta_{PG}$: pseudogap; $\Delta_{SC}$: superconducting gap) with temperature. Spectra are offset for clarity in (e). (c,f) Blow up of the superconducting gap energy scale showing the destruction of the superconducting gap in a magnetic field of $H$ = 5.7 T > $H_{c2}$ while the pseudogap feature is preserved. The spectra *G(V)* in (c) and (d) are normalized by their corresponding junction impedances $G_S$.

**Fig.2: Quasiparticle interference of heavy superconducting electrons**. Real space conductance map (a) and its DFT (b) at a bias of 1.5 mV measured at $T$ = 245 mK on surface B. Colorbar in (a) denotes deviation from the mean. $Q_1$, $Q_2$, $Q_3$ correspond to the different quasiparticle scattering vectors. (c) DFT at $V$ = 3 mV (see also supplementary section III). Axes in (c) denote the Bragg orientation for all DFTs and for the schematic (e). (d) Energy-momentum structure of $Q_1$, $Q_2$, $Q_3$ showing rapid dispersions reflective of mass enhancements $m^*$ = 34 $m_0$, 29 $m_0$, 23 $m_0$ respectively. Error bars are derived from the width of the peaks in the DFTs. (e) Schematic of the band structure in the first Brillouin zone derived from Refs. 21-23 showing the $\alpha$ (magenta), $\beta$ (blue) and small (orange) Fermi surfaces in the $k_z$ = 0 (solid) and $k_z$ = $\pi$ (dashed) planes. The measured $Q_1$, $Q_2$ & $Q_3$ QPI scattering vectors are drawn to scale for comparison (see also supplementary section IV). DFTs for selected energies in the superconducting (f-j) and normal

(k-o) states. The Q-space range of the DFTs in (b,c,f-o) is ±0.5 rlu, where 1 rlu = $2\pi/a_0 = 2\pi/4.6$ Å.

Fig.3: **Evolution of in-gap quasiparticle states approaching a step-edge.** (a) Topographic image ($V$ = -100 mV, $I$ = 100 pA) of surface A showing a single unit-cell step-edge oriented at 45° to the atomic lattice. The arrows in the figure indicate the in-plane crystallographic **a** and **b** directions (b) Evolution of the spectra near the step-edge: $G(V)$ subtracted by the spectrum far away from the step-edge $G(V, r = 153$ Å$)$. The locations of the spectra in (b) are plotted on (a). (c) Schematic representation of nodal superconducting quasiparticles scattering off a step-edge. (d) Zero-bias conductance $G_0(r)$ subtracted by the extrapolated $G_0(r = \infty)$ as a function of distance from the step edge. Line represents an exponential fit to the data, where error bars denote the standard deviation on the averaged spectra. $\xi_{BCS}$ denotes the characteristic decay length obtained from the fit in (d), which is a measure of the BCS coherence length.

Fig.4: **Visualizing the vortex lattice and vortex-bound quasiparticle states.** Zero-bias conductance maps both taken at $H$ = 1 T (separate field dials) and at $T$ = 245 mK show the vortex lattice structure expected below (a) and above (b) the transition seen at this field by neutron scattering in Ref. 29. The arrows in the figure indicate the in-plane crystallographic **a** and **b** directions (c) Close-up zero-bias map of the vortex lattice on surface B showing an anisotropic square vortex core ($H$ = 1.5 T). (d) Line-cut of spectra starting from the center of a vortex and moving radially outward at 45 degrees to the **b**-axis showing the evolution of the bound state inside the superconducting gap ($H$ = 0.5 T). (e) Radial dependence of the angularly averaged zero-bias conductance $G_0$ for a single vortex core at $H$ = 1 T. Error bars (estimated from the standard deviation in the analyzed map) are smaller than the marker size in (e). Inset shows the angular dependence of the radially averaged conductance showing the four-fold anisotropy of a single vortex

with higher conductance extending along the **a**- and **b**-directions directions (see supplementary section VIII). $\xi_{GL}$ denotes the characteristic decay length obtained from the fit in (e), which is a measure of the angularly averaged Ginzburg-Landau coherence length.

**Fig.5: Visualizing impurity-bound quasiparticle excitations.** (a) Topographic image of an impurity on surface B ($V$ = -6 mV, $I$ = 100 pA). (b) Model calculation for the real space structure (roughly 10 Fermi wavelengths across) of the hole-like part of the impurity bound state in a $d_{x2-y2}$ superconductor, reproduced from Ref. 32 (Copyright (2000) by the American Physical Society). (c) Electron-like state for the same impurity in (b). (d-g) Local density of states obtained on the same field-of-view as (a) at ± 195 µV in the normal ($H > H_{c2}$) and superconducting ($H = 0$) states as indicated on the figure. Colorbar in (d-g) denotes deviation from the mean. (h) Radial average of the density of states across the lobes measured in (d,e), normalized to their sum, as a function of angle from the **a** axis. Data at negative (positive) energy is shown in blue (red) symbols; the lines are guides to the eye. A $d_{x2-y2}$ gap is shown in yellow.

# Figure 1

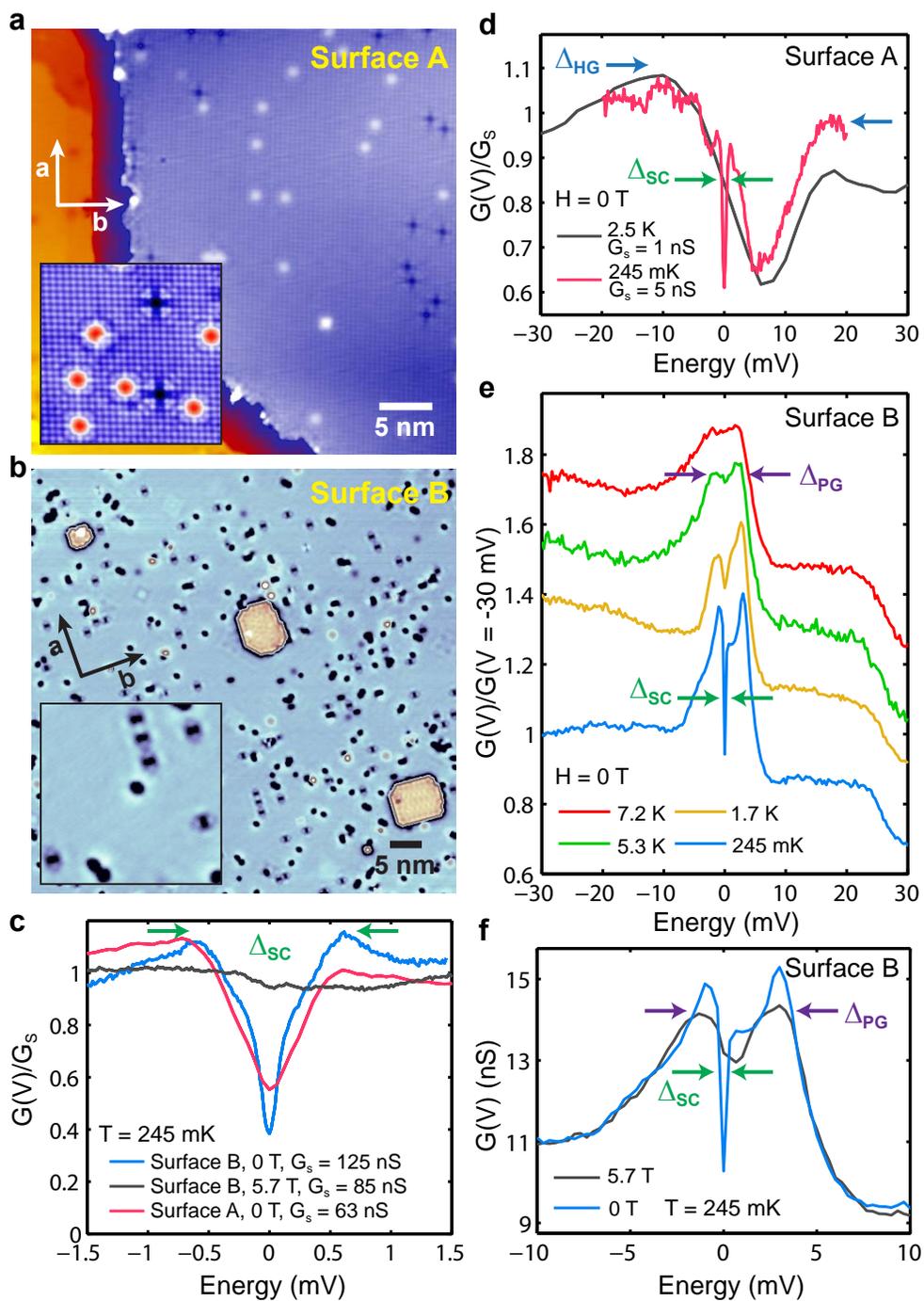

**Figure 2**

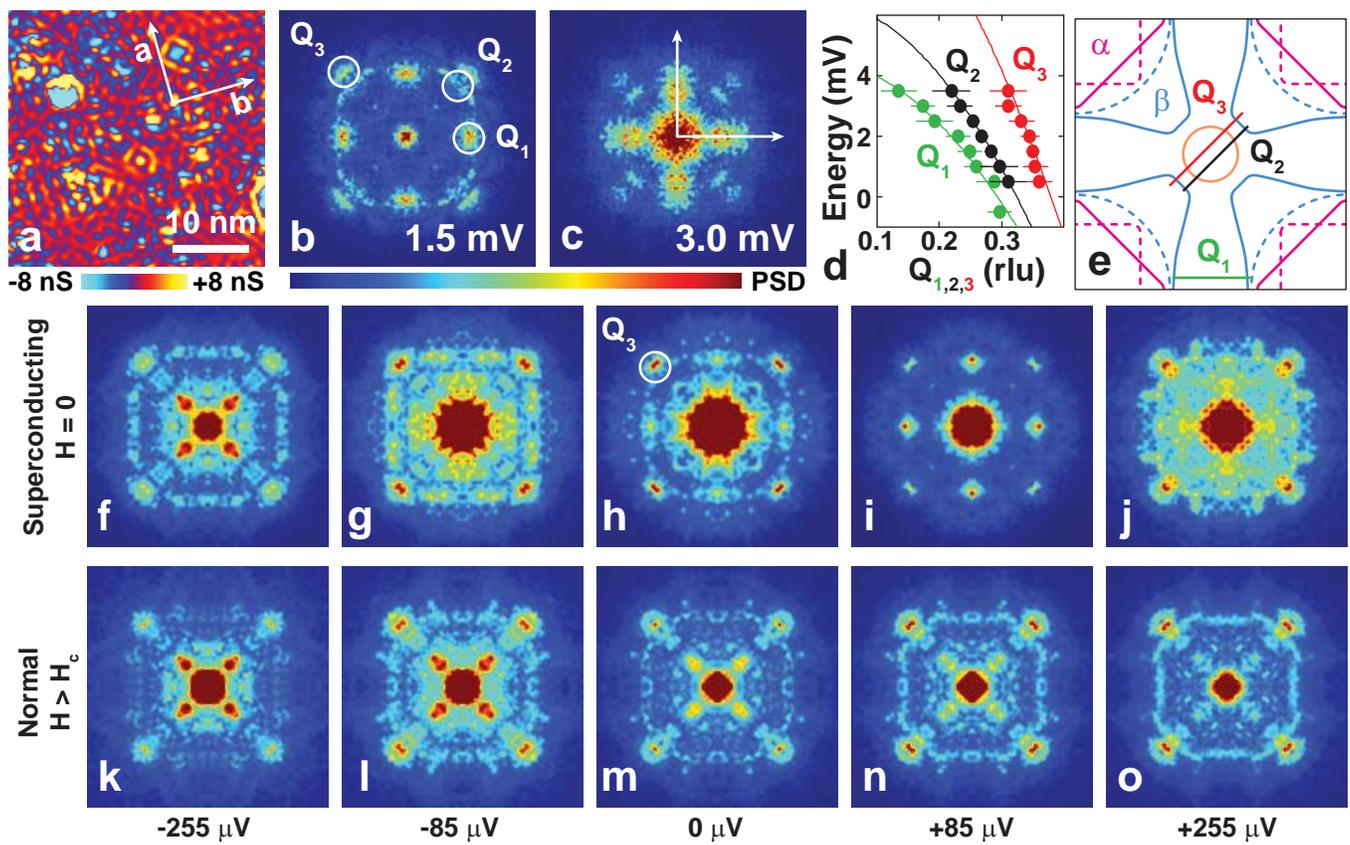

**Figure 3**

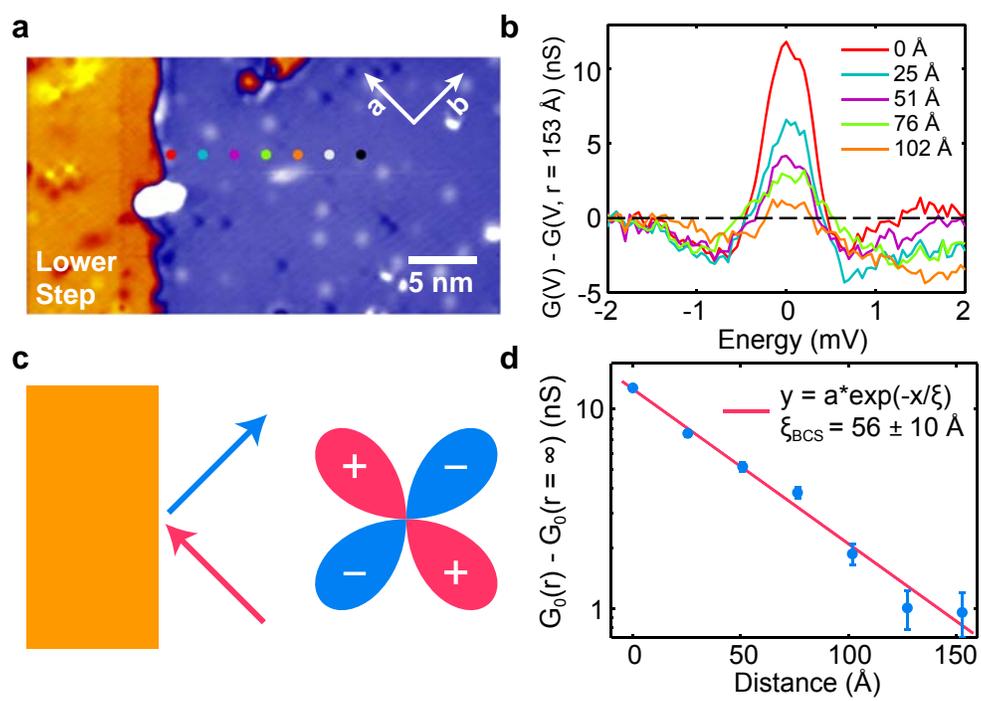

**Figure 4**

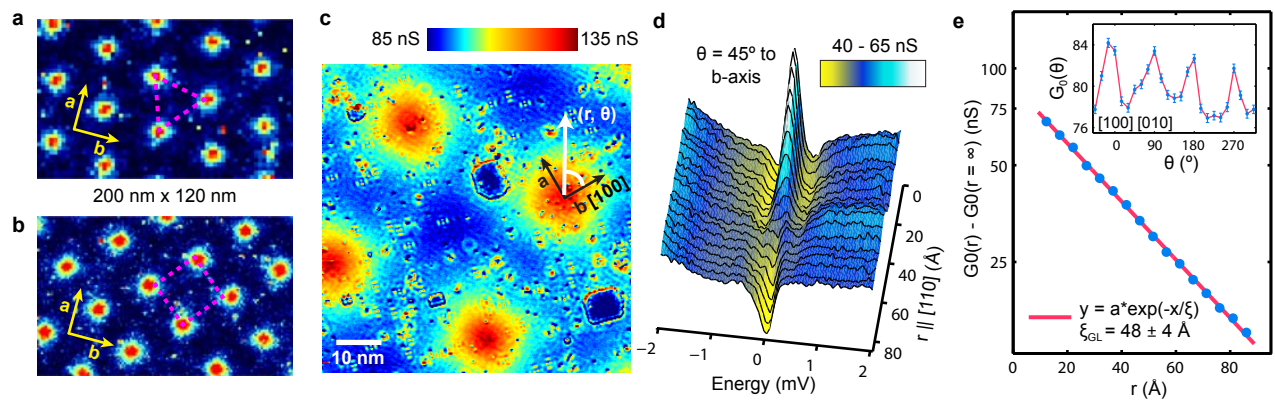

**Figure 5**

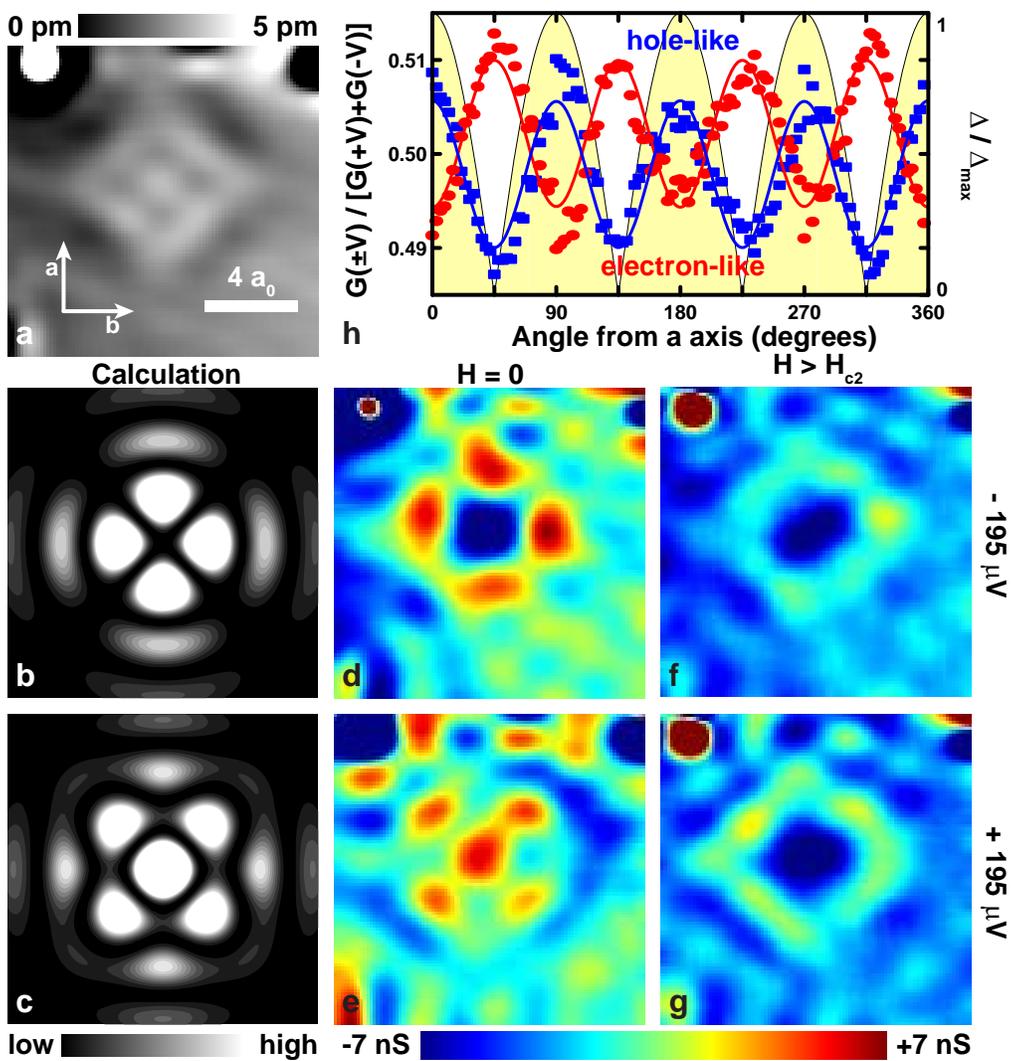

# Supplementary Information for "Visualizing Nodal Heavy Fermion Superconductivity in CeCoIn$_5$"

Brian B. Zhou[1*], Shashank Misra[1*], Eduardo H. da Silva Neto[1], Pegor Aynajian[1], Ryan E. Baumbach[2], J. D. Thompson[2], Eric D. Bauer[2], and Ali Yazdani[1*]

[1] Joseph Henry Laboratories & Department of Physics, Princeton University, Princeton, New Jersey 08544, USA
[2] Condensed Matter and Magnet Science, Los Alamos National Laboratory, Los Alamos, New Mexico 87545, USA
* These authors contributed equally to this work.

## SI. 1$^{st}$ order phase transition at the upper critical field

Fig. S1a shows spectra corresponding to the local electronic density of states measured on surface A at $T$ = 245 mK as a function of magnetic field across the upper critical field $H_{c2}$ = 5 T. For fields below the transition, a conductance map was taken to locate the position of the vortices, and the spectrum is then measured on a location away from a vortex. The spectra show a sudden shift in their zero bias conductances at $H_{c2}$, resembling a 1$^{st}$ order phase transition, as seen in bulk transport measurements in CeCoIn$_5$ (S1,S2). Fig. S1b shows the zero bias conductance $G(V = 0)$ as a function of magnetic field across $H_{c2}$.

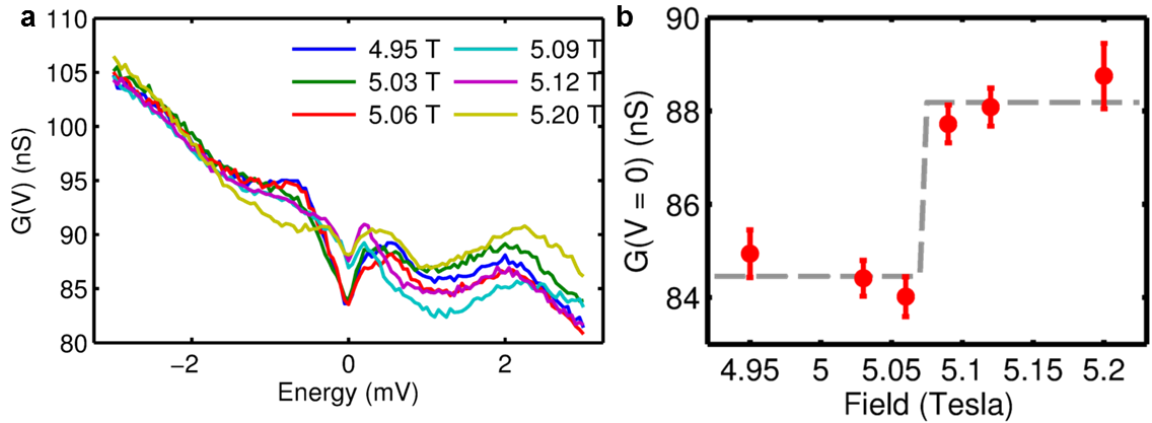

**Fig. S1: Superconducting phase transition at the upper critical field H$_{c2}$.**

## SII. Measurements of the superconducting gap in single-crystal Al upon approaching an atomic step edge.

Fig. S2a shows an STM topographic image with atomic step edges obtained on the surface of a single-crystal Al ($T_c$ = 1.2 K). Fig. S2b display spectra corresponding to the local density of states showing the superconducting energy gap across the step edge (locations indicated by red dots in (a)). The gaps show no sensitivity to the proximity to the step edge, consistent with the *s*-wave nature of the superconducting gap, which is robust against potential scattering. Fig. S2c shows the superconducting gap and the corresponding thermally broadened BCS fit, indicating an electronic temperature of 245 ± 20 mK in our low temperature STM.

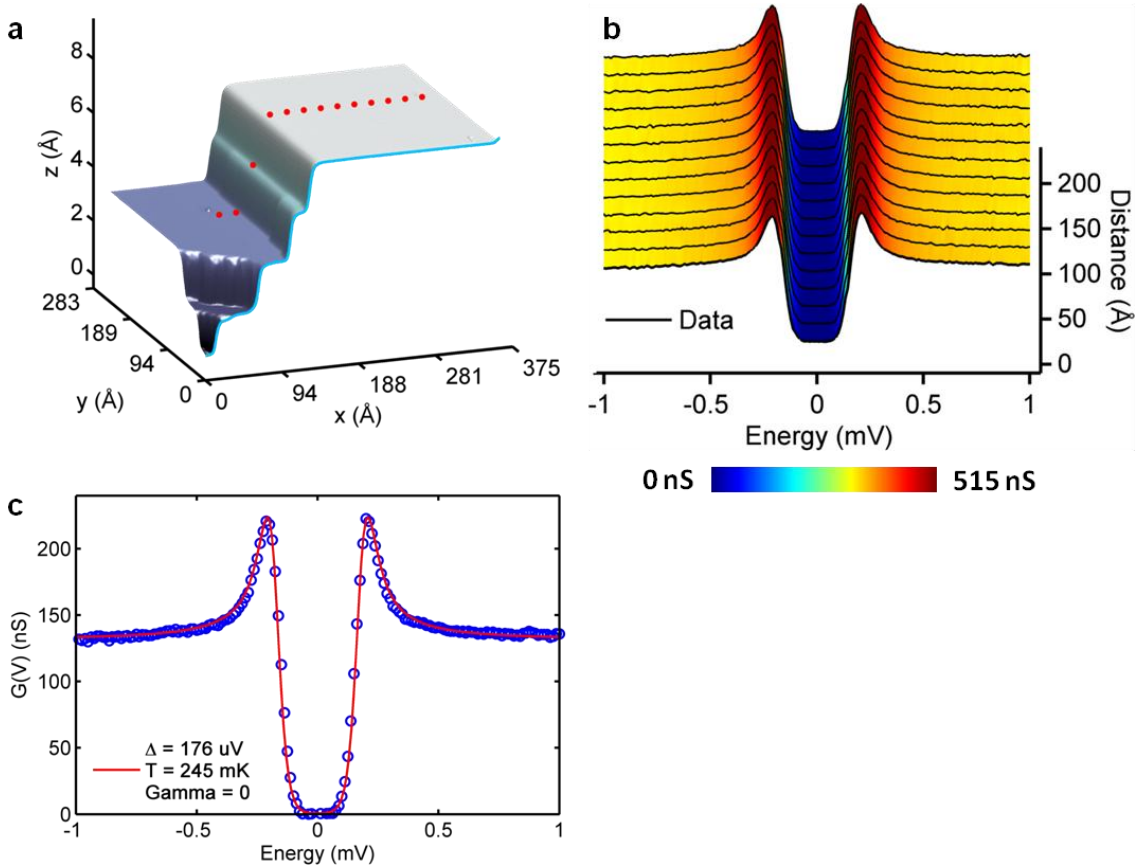

**Fig. S2: Superconducting gap in Al across an atomic step edge.**

## SIII. Energy-momentum structure of the heavy quasiparticle band.

Fig. S3 shows the topograph and the corresponding real space conductance map at 1.5 mV with a set-point current of 100 pA and bias of -6 mV taken on surface B over a field of view of 67 nm. All quasiparticle interference (QPI) data were carried out on this entire field of few (in Fig. 2a of the main manuscript we zoom in on only a quarter of the real space field of view). The small islands in the topograph correspond to portions of the residual top layer.

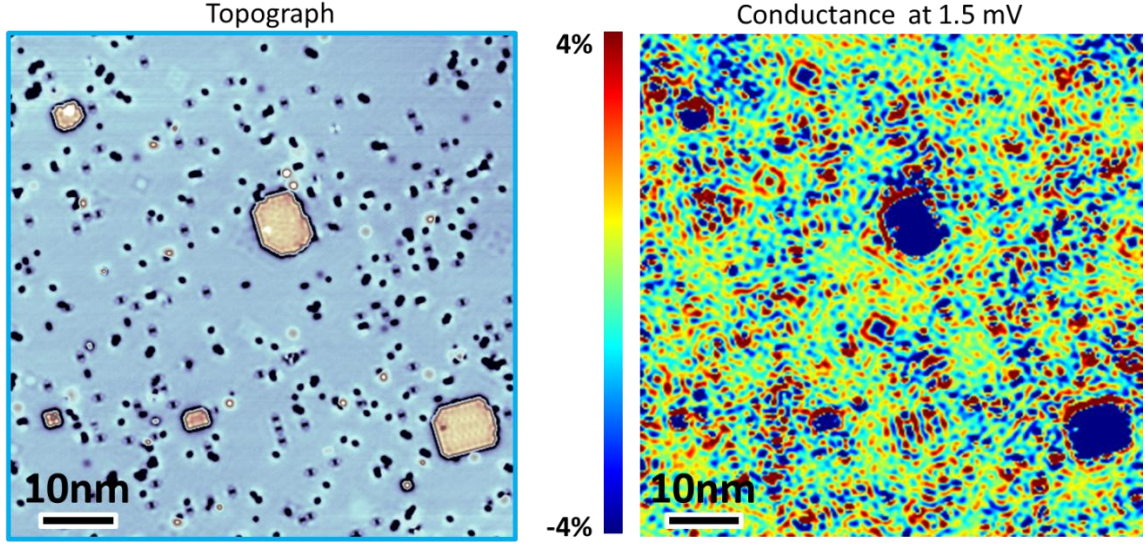

**Fig. S3: Topograph and conductance map (normalized to the mean) on surface B.**

Fig. S4 shows the discrete Fourier transforms of the conductance maps over the same field of view as Fig. S3 for selected energies near the Fermi energy. We identify three distinct features: $Q_1$ along $(\pi, 0)$ and $Q_{2,3}$ along $(\pi, \pi)$. As is evident from Fig. S6a,b, all three features shorten rapidly with increasing bias, manifesting heavy quasiparticle effective masses between 20-40 $m_0$. For comparison to previous experimental knowledge on the band structure of $CeCoIn_5$, we note that at the Fermi energy, $Q_1$ = (0.29,0) rlu, $Q_2$ = (0.24,0.24) rlu, and $Q_3$ = (0.27,0.27) rlu, where statistical and systematic uncertainties total ± 0.03 rlu, where 1 rlu = $2\pi/a_0$ = $2\pi/(0.46$ nm).

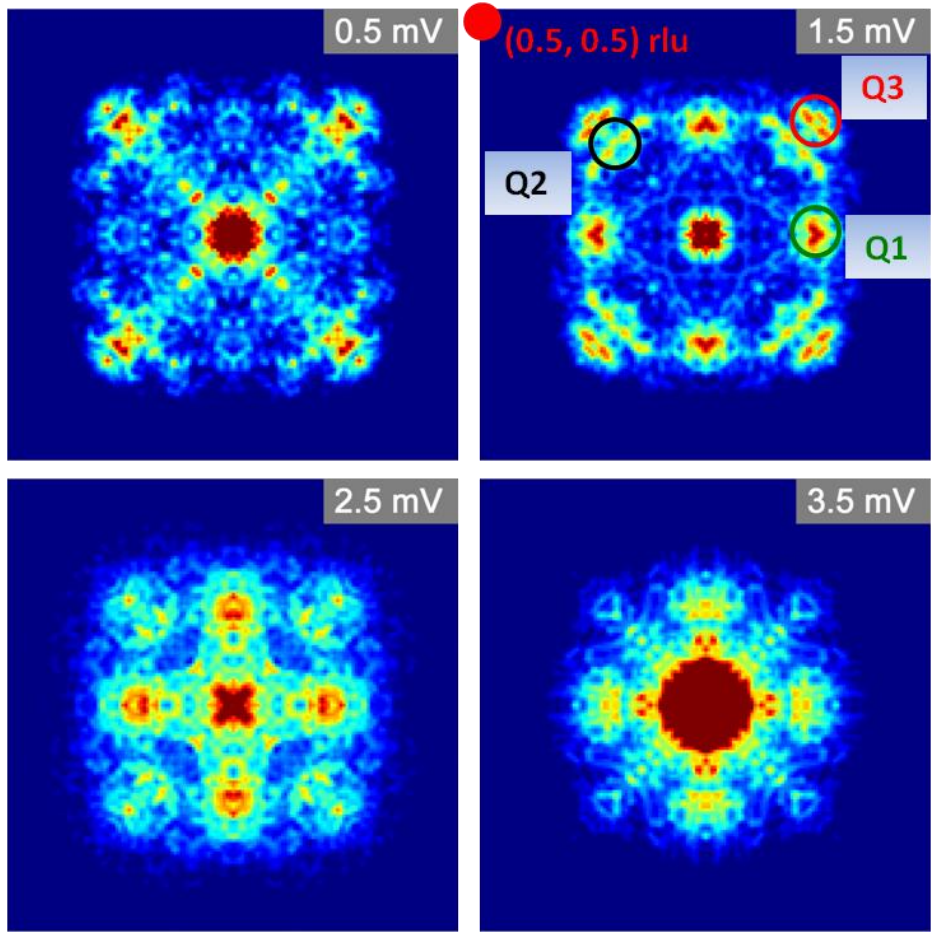

**Fig. S4: Dispersion of the heavy quasiparticle bands on surface B.**

### SIV. The Fermi surface inferred from QPI in context of current understanding of CeCoIn$_5$ band structure

Since QPI measures the momentum transfer vectors (*Q*) which connect the Fermi surface (FS), rather than the *k*-vectors of the FS directly, inferring a unique FS from QPI in a three-dimensional, multi-band material without making a large number of assumptions is not possible. As schematically illustrated in Fig. S5 (a), three bands cross the Fermi level of CeCoIn$_5$ (identified previously by various theoretical and experimental efforts, see main text): band 135 (i.e., "α band" with cylindrical Fermi surfaces around the M points), band 133 (i.e., "β band" with a large, complicated Fermi surface), and finally band 131 with small Fermi surfaces (S3,S4). Because of the expected light mass for band 131, inconsistent with the rapid dispersions seen in our Q vectors, and its small size, we focus instead on the α and β bands for the origin of our Q vectors (specifically to only the surfaces of α and β seen in quantum oscillation experiments). The decrease in length of our Q vectors with increasing energy restricts us to look for scattering between two disconnected surfaces of the α and β bands rather than for scattering within a single closed surface. By the same dispersion argument, interband scattering between concentric α and β sheets can also be excluded as a possibility since in general β disperses faster with increasing energy than α does (lengthening the Q with increasing energy). In the table of Fig S5b, we convert the measured de Haas-van Alphen (dHvA) frequencies to Fermi surface areas in units of 1 Brillouin zone. The extremal areas of each band then give lower limits on the length of possible connecting Q vectors, which we estimate by assuming simple FS shapes consistent with theory. It is immediately apparent that the $(\pi,\pi)$ Q vectors ($Q_{2,3}$) can only come from the β band as the α band cylinders are too far separated in that direction. However, $Q_1$ along $(\pi,0)$ may originate from either the α or β band as the measured $Q_1$ = (0.29,0) can originate on the zone edge where the two bands are close together.

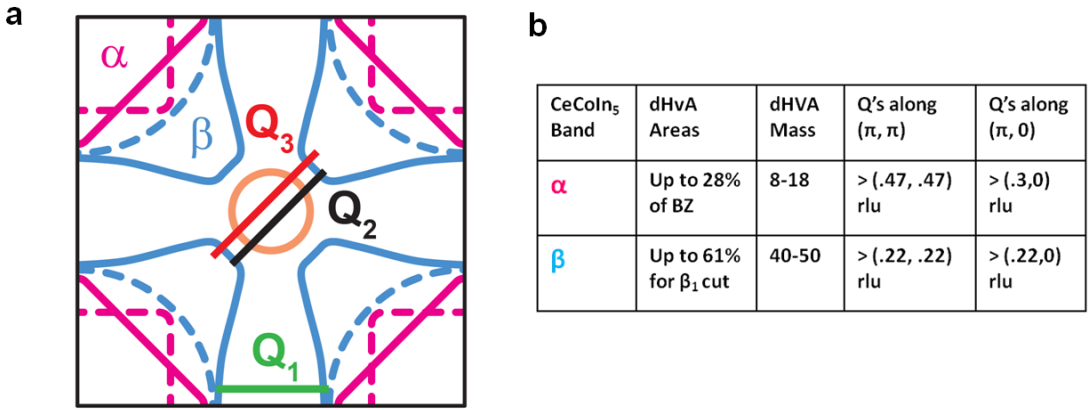

**Fig. S5: Comparison to Fermi surfaces of CeCoIn5 deduced from dHvA.** (a) FS cuts perpendicular to [001] for band 135 ("α band", magenta), band 133 ("β band", cyan), and band 131 (orange) derived from ref. S3. The measured Q vectors of the QPI are drawn to scale on top. (b) Maximal dHVA areas perpendicular to [001] as a percentage of the first Brillouin zone and cyclotron mass from ref. S3. Corresponding to each extremal area, we estimate the vector of closest approach in the two high-symmetry directions. This analysis indicates that $Q_{2,3}$ should originate from the heavy β band, while $Q_1$ can originate from either α or β (not necessarily from the exact location drawn in (a)).

## SV. Phenomenological Modeling of Normal State Band Structure

To speculate on the qualitative features of QPI in the superconducting state, we first capture the energy dispersions of the normal state in an over-simplified 2-dimensional (2D) model. For concreteness, we identify two 2D surfaces (corresponding for example to two different $k_z$ cuts of the $\beta$ band) whose energy dispersions $\varepsilon_k$ and $\chi_k$ are given by

$$\varepsilon(k_x, k_y) = \mu + t_1\big(\cos(k_x) + \cos(k_y)\big) + t_2 \cos(k_x)\cos(k_y) + t_3\big(\cos(2k_x) + \cos(2k_y)\big)$$
$$\chi(k_x, k_y) = v + s_1\big(\cos(k_x) + \cos(k_y)\big) + s_2 \cos(k_x)\cos(k_y) + s_3\big(\cos(2k_x) + \cos(2k_y)\big)$$
$$(\mu, t_1, t_2, t_3) = (-31.3, 67.2, -124.6, 13.2) \text{ mV}$$

$(v, s_1, s_2, s_3) = (-54.6, 36.3, 5.0, -9.2)$ mV.

By suitable adjustment of the hopping parameters, the $2k_f$ scatterings within this model can be made to reproduce the dispersions and general QPI pattern measured in the experiment as shown in Fig. S6.

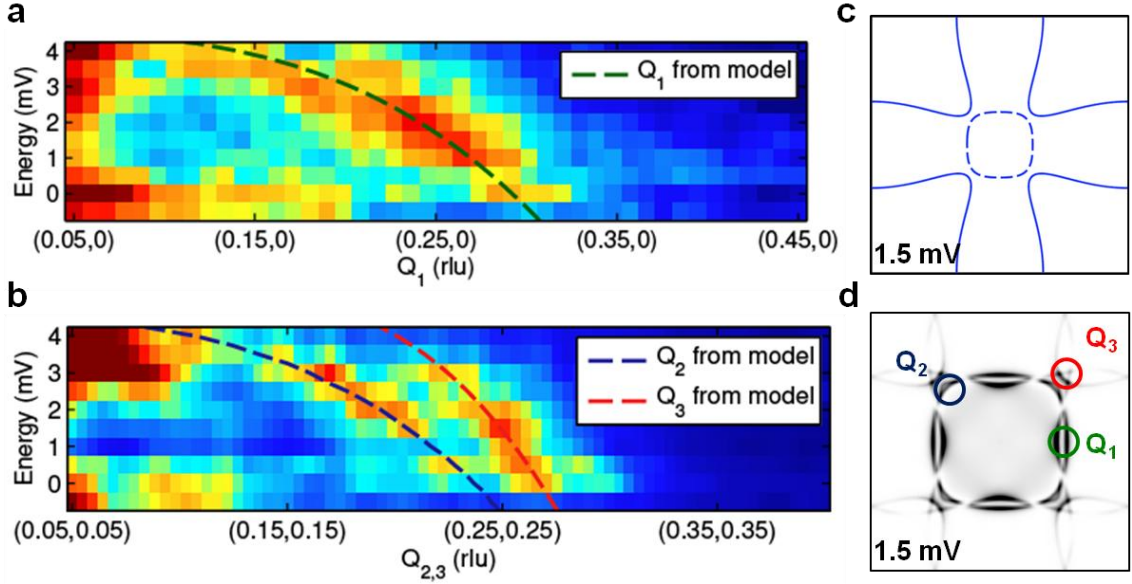

**Fig S6: Normal State QPI Model.** (a,b) show the experimental QPI peaks dispersing along $(\pi, 0)$ and $(\pi, \pi)$ directions, respectively, overlaid with the dispersion of the appropriate $2k_f$ scatterings calculated from the parameterized 2D $\varepsilon(k)$ and $\chi(k)$ surfaces. The in-field H = 5.7 T data was substituted for the linecut at 0 energy. (c) shows the typical constant energy contour for $\varepsilon(k)$ (blue solid) and $\chi(k)$ (blue dashed). The simulated Born scattering QPI pattern for 1.5 mV is shown in (d).

To calculate the QPI patterns, we applied the Born scattering approximation:

$$S(q, \omega) = \frac{dI(q, \omega)}{dV} = \frac{2\pi e}{\hbar} N_t \sum_{i,j=1}^{2} [\hat{t}\ \hat{N}(q, \omega)\ \hat{t}]_{i,j}$$

$$\hat{N}(q, \omega) = -\frac{1}{\pi} Im \int \frac{d^2 k}{(2\pi)^2} \hat{G}(k, \omega) \hat{U}\ \hat{G}(k+q, \omega)$$

where $S(q, \omega)$ reflects experimental Fourier transform of the differential conductance (S5,S6). We take the density of states of the STM tip $N_t = 1$, $\hat{t} = \begin{pmatrix} t_\varepsilon & 0 \\ 0 & t_\chi \end{pmatrix}$ with $t_\varepsilon = -1$ and $t_\chi = 0.7$ denoting the propensity to tunneling into the $\varepsilon$ and $\chi$ surfaces, and $\hat{G} = \begin{pmatrix} G_\varepsilon^0 & 0 \\ 0 & G_\chi^0 \end{pmatrix}$ encoding the full Green's functions $G_\varepsilon^0(k, \omega) = (\omega + i\Gamma_\varepsilon - \varepsilon(k))^{-1}$ and $G_\chi^0(k, \omega) = (\omega + i\Gamma_\chi - \chi(k))^{-1}$.

The lifetimes $\Gamma_\varepsilon$ and $\Gamma_\chi$ are taken to be 0.1 mV for the normal state, and the scattering matrix $\widehat{U} = \begin{pmatrix} U_\varepsilon & U_{\varepsilon\chi} \\ U_{\varepsilon\chi} & U_\chi \end{pmatrix} = \begin{pmatrix} 1 & .3 \\ .3 & .7 \end{pmatrix}$.

Finally, we note that the broad feature $Q_1$ along $(\pi, 0)$ seen in experiment may come from an overlap of intra-surface scattering from both the $\varepsilon_k$ and $\chi_k$ surfaces, and would be more precisely captured in a full 3D model that considers $k_z$ dispersion.

## SVI. Superconductivity Gapping the Phenomenological Band Structure

We investigate how superconductivity qualitatively changes the QPI patterns by applying both a $d_{x^2-y^2}$ and $d_{xy}$ gap function on our model normal state band structure $\varepsilon_k$ and $\chi_k$. In the presence of superconductivity, the Green's functions for $\varepsilon_k$ and $\chi_k$ acquire particle/hole channels given by the 2x2 matrices $G_\varepsilon^S(k,\omega) = \left((\omega + i\Gamma_\varepsilon)\,I - \varepsilon(k)\sigma_3 - \Delta(k)\sigma_1\right)^{-1}$ and $G_\chi^S(k,\omega) = \left((\omega + i\Gamma_\chi)\,I - \chi(k)\sigma_3 - \Delta(k)\sigma_1\right)^{-1}$, where $\sigma_i$ are the Pauli matrices. The above equations for $S(q,\omega)$ and $\widehat{N}(q,\omega)$ still hold with now

$$\widehat{G} = \begin{pmatrix} G_{\varepsilon\,11}^S & 0 & G_{\varepsilon\,12}^S & 0 \\ 0 & G_{\chi\,11}^S & 0 & G_{\chi\,12}^S \\ G_{\varepsilon\,21}^S & 0 & G_{\varepsilon\,22}^S & 0 \\ 0 & G_{\chi\,21}^S & 0 & G_{\chi\,22}^S \end{pmatrix}, \quad \hat{t} = \begin{pmatrix} -t_\varepsilon & 0 & 0 & 0 \\ 0 & -t_\chi & 0 & 0 \\ 0 & 0 & t_\varepsilon & 0 \\ 0 & 0 & 0 & t_\chi \end{pmatrix}$$

and considering potential scattering

$$\widehat{U} = \begin{pmatrix} U_\varepsilon & U_{\varepsilon\chi} & 0 & 0 \\ U_{\varepsilon\chi} & U_\chi & 0 & 0 \\ 0 & 0 & -U_\varepsilon & -U_{\varepsilon\chi} \\ 0 & 0 & -U_{\varepsilon\chi} & -U_\chi \end{pmatrix}.$$

Using identical model parameters as in the normal state calculation (with the exception of $\Gamma_\varepsilon = \Gamma_\chi = 0.05$ mV), we simulate the experimental QPI at three energies for a $d_{x^2-y^2}$ gap on the $\varepsilon_k$ and $\chi_k$ surfaces (Fig. S7b)

$$\Delta^{\varepsilon,\chi}(k_x, k_y) = \frac{\Delta_{x^2-y^2}^{\varepsilon,\chi}}{2}(\cos(k_x) - \cos(k_y))$$

with $\Delta_{x^2-y^2}^\varepsilon = 0.67$ mV and $\Delta_{x^2-y^2}^\chi = 2.55$ mV such that the maximum gaps on $\varepsilon$ and $\chi$ are 0.5 mV. The assumption of equal gaps on both $\varepsilon$ and $\chi$ is an arbitrary feature of our model. Since our high resolution QPI data show subtle changes between the superconducting and normal states throughout Q-space (see Fig. 2 of main text), no unambiguous identification of completely ungapped portions of the FS can be made; likewise, determining the size of the gap on different surfaces based on these subtle features would be purely speculative. Alternatively, we can also consider a $d_{xy}$ gap (Fig. S7 c)

$$\Delta^{\varepsilon,\chi}(k_x, k_y) = \Delta^{\varepsilon,\chi}_{xy}(\sin(k_x) * \sin(k_y))$$

with $\Delta^{\varepsilon}_{xy} = 0.67$ mV and $\Delta^{\chi}_{xy} = 1.09$ mV, maintaining a maximal 0.5 mV gap on both surfaces. Comparison of the panels shows that the experimental data cannot be reconciled with a $d_{xy}$ gap on the $\varepsilon$ surface, and is qualitatively most consistent with a $d_{x^2-y^2}$ gap on both surfaces. However, such analysis cannot reproduce the strong electron-hole asymmetry displayed by the data in Fig. 2f-o of the main text, whose explanation may require additional assumptions about impurity effects or ungapped regions of the Fermi surface. These assumptions, absent independent experimental justification, together with the complex 3D, multi-band Fermi surface of CeCoIn$_5$, make extraction of the superconducting gap from QPI data ambiguous.

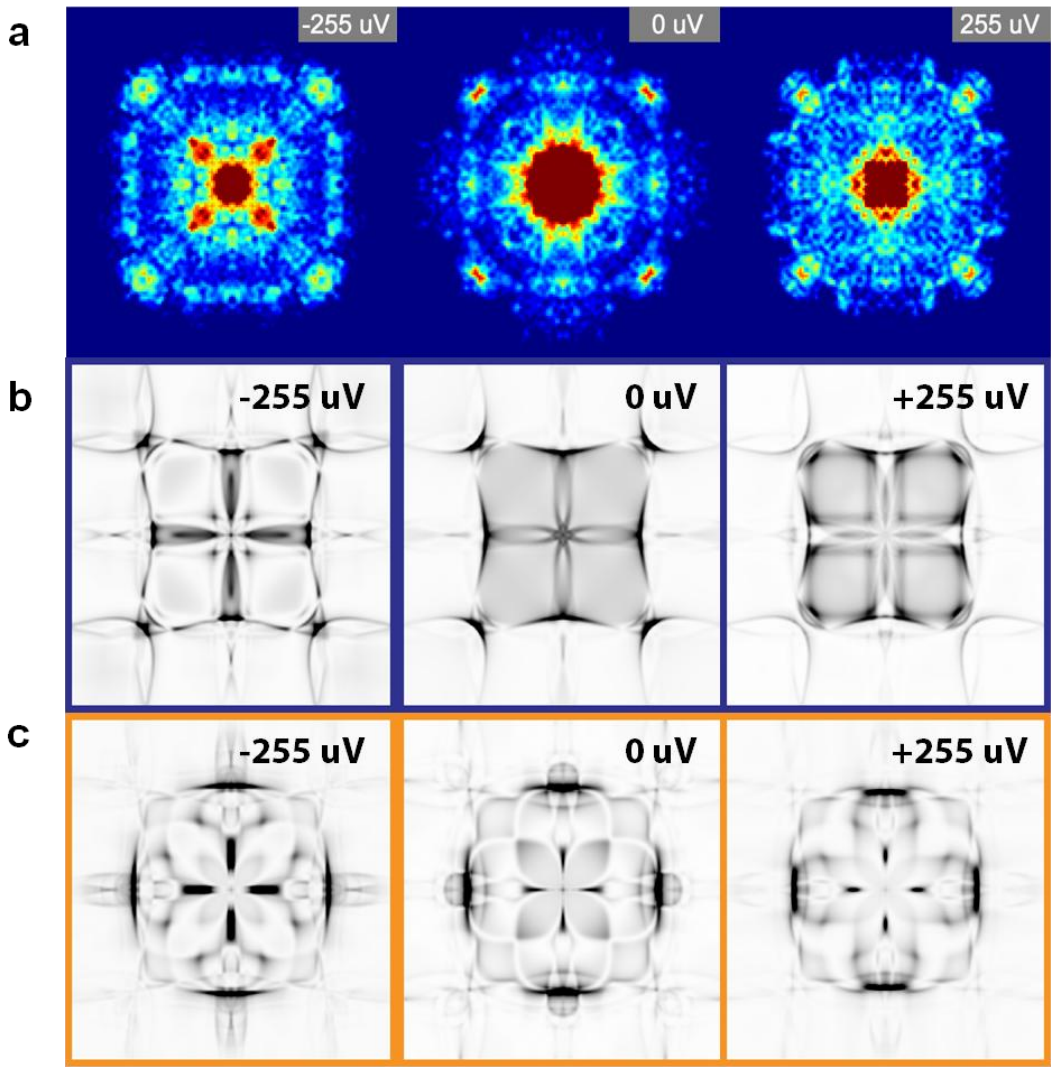

**Fig S7: Superconducting QPI in comparison to simulation of $d_{x^2-y^2}$ and $d_{xy}$ gap symmetries.** (a) shows the experimental data, while the bottom two rows show the resulting QPI pattern with application of a $d_{x^2-y^2}$ gap (b) and $d_{xy}$ gap (c) on our phenomenological band structure model.

## SVII. Zero Bias Enhancement of Nodal QPI Peak Amplitude

In Fig S8a, we plot the amplitude of the power spectral density (PSD) of the $(\pi, \pi)$ $Q_3$ feature identified in the QPI for both the normal and superconducting states. $Q_3$ was identified in Section IV as originating from the β band. Error bars represent the $1*\sigma$ error in the amplitude parameter obtained from fitting the unnormalized PSD around $Q_3$ to a Gaussian. We note that while the overall differential conductance at zero bias is reduced due to the superconducting gap, the QPI strength, in contrast, is enhanced. Subtraction of the peak PSD of the $Q_3$ feature in the normal state from the superconducting state shows enhancement centered at zero energy, consistent with the confinement of the Fermi surface around the $(\pi, \pi)$ nodes by a $d_{x^2-y^2}$ superconducting gap (Fig S8b).

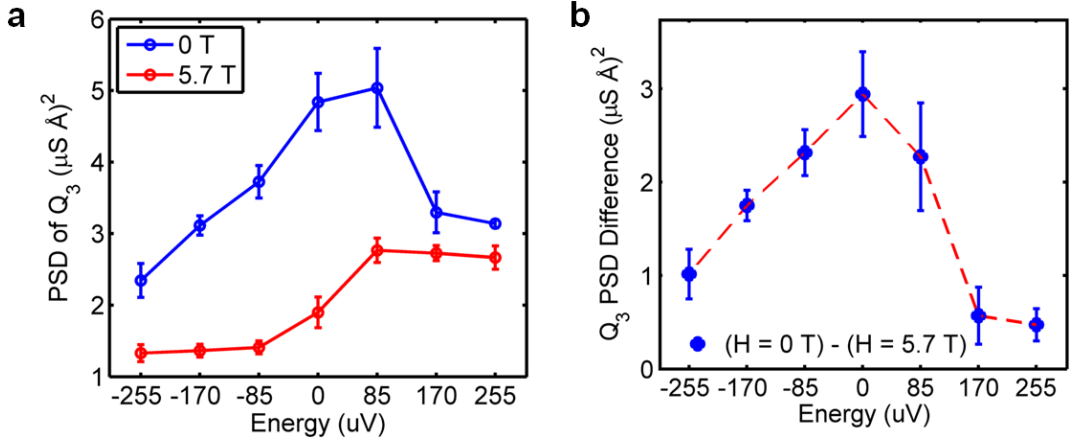

**Fig S8: Power Spectral Density of $Q_3$.**

## SVIII. Four-fold anisotropy of the vortex bound state in CeCoIn$_5$

Fig. S9a shows the zero-bias conductance map of the single vortex core analyzed in Fig. 4e of the main text. Data was obtained with a set-point current of 300 pA and set-point bias of -3 mV at $H$ = 1 T on surface A of CeCoIn$_5$. The radial analysis was performed by taking the average of the conductance in annular rings ($\Delta r$ = 4.9 Å) between the two radii shown, while the angular analysis used angular wedges ($\Delta\theta$ = 15°) between the two radii (main part of Fig. 4e and inset, respectively). In Fig. S9b we measure spectra along the two high symmetry axes of the vortex core with a set-point current of 100 pA and set-point bias of -2 mV at $H$ = 0.5 T. The spectra show the bound state within the superconducting energy-gap scale to extend further out along the atomic ***a*** or ***b*** directions ($\theta$ = 0° or [100]) compared to that along the diagonal direction ($\theta$ = 45° or [110]).

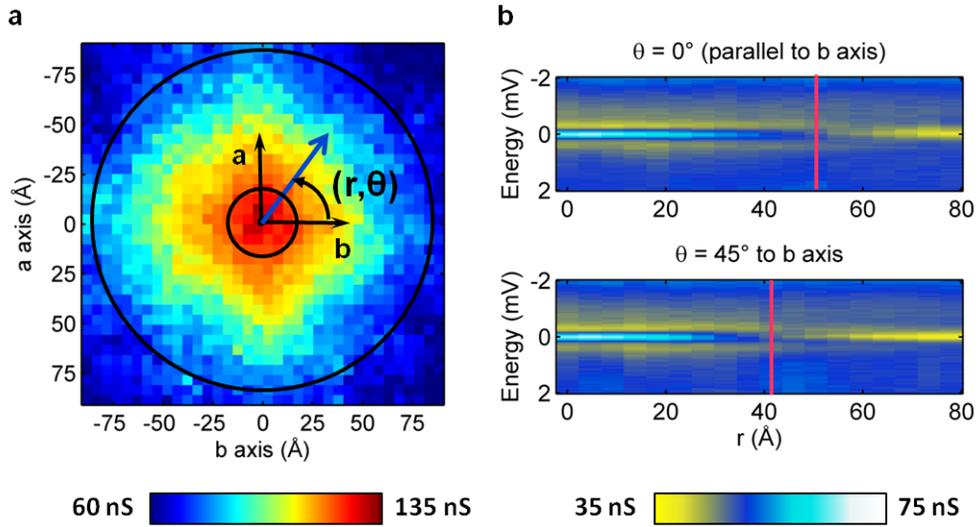

**Fig.S9: Anisotropy of the vortex bound state.**

## SIX. Spatially complementary in-gap density of states induced by the impurity

Fig. S10a shows spectra measured at H = 0 T for various locations near the impurity shown in Fig. 5 of the main text. The colored dots in the conductance maps shown in Figs. S10b,c (reproduced from the main text) indicate the location of the spectra, which are all normalized to the spectrum taken away from the impurity (black dot in Fig. S10b). There is an enhancement of the in-gap density of states at negative energies in the spectrum taken at one of the lobes of the hole-like conductance map (Figs. S10a,b, blue). Complementary to that, there is an enhancement of the in-gap density of states at positive energies in the spectrum taken at one of the lobes of the electron-like conductance map (Figs. S10a,c, red). The spectrum taken at the center of the impurity (Figs. S10a,c, green) shows that the on-site impurity resonance occurs at positive energy.

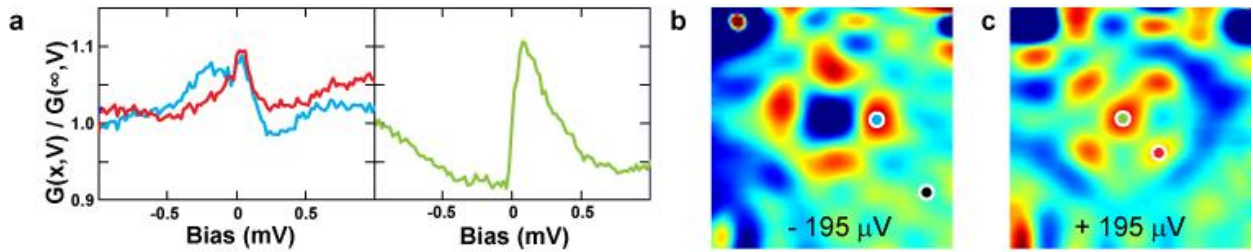

**Fig. S10: Impurity bound state**